\documentclass[notitlepage,aps,twocolumn,superscriptaddress,longbibliography,nofootinbib]{revtex4-1}
\usepackage{amsmath}
\usepackage{amssymb}
\usepackage{todonotes}
\usepackage[unicode=true,colorlinks=true,citecolor=blue,urlcolor=blue]{hyperref}
\usepackage{float}
\usepackage{bm}
\usepackage{epsfig}
\usepackage{lipsum}
\usepackage{xcolor,soul}
\usepackage[normalem]{ulem}

\renewcommand {\Im}{\mathop\mathrm{Im}\nolimits}
\renewcommand {\Re}{\mathop\mathrm{Re}\nolimits}

\renewcommand {\i}{{\rm i}}
\renewcommand {\phi}{{\varphi}}

\begin{document}
\title{Numerical and theoretical study of eigenenergy braids in two-dimensional photonic crystals}

\author{Janet Zhong}
\affiliation{Department of Applied Physics, Stanford University, Stanford, California 94305, USA}
\author{Charles C. Wojcik}
\affiliation{Department of Electrical Engineering, Ginzton Laboratory, Stanford University, Stanford, California 94305, USA}
\author{Dali Cheng}
\affiliation{Department of Electrical Engineering, Ginzton Laboratory, Stanford University, Stanford, California 94305, USA}
\author{Shanhui Fan}
\email{shanhui@stanford.edu}
\affiliation{Department of Applied Physics, Stanford University, Stanford, California 94305, USA}
\affiliation{Department of Electrical Engineering, Ginzton Laboratory, Stanford University, Stanford, California 94305, USA}

\begin{abstract}
We consider non-Hermitian energy band theory in two-dimensional systems, and study eigenenergy braids on slices in the two-dimensional Brillouin zone. We show the consequences of reciprocity and geometric symmetry on such eigenenergy braids. The point-gap topology of the energy bands can be found from the projection of the eigenenergy braid onto the complex energy plane. We show that the conjugacy class transitions in the eigenenergy braid results in the changes in the number of bands in a complete point-gap loop. This transition occurs at exceptional points. We numerically demonstrate these concepts using two-dimensional reciprocal and nonreciprocal photonic crystals.
\end{abstract}
\date{\today}

\maketitle
\section{Introduction}

A distinct feature of non-Hermitian band theory is the eigenenergy braiding, where the complex eigenenergies form braids as the momentum varies along a one-dimensional (1D) loop trajectory~\cite{bergholtz2021exceptional,ding2022nonhermitian}. For 1D systems, the eigenenergy topology of separable energy bands is completely classified by the braid group that describes these braids as the momentum varies across the 1D Brillouin zone~\cite{wojcik2020homotopy,hu2021knots,li2021homotopical}. For two-dimensional (2D) systems, the eigenenergy topology can also be classified by studying the relationships between the braids along different 1D loops in the 2D Brillouin zone~\cite{wojcik2022eigenvalue,hu2022knot,konig2022braid,guo2022exceptional}.

Eigenenergy braiding is intimately connected to the study of exceptional points. Exceptional points are points of degeneracies where both the eigenenergy and eigenvectors coalesce~\cite{heiss2012physics,miri2019exceptional}. They are unique to non-Hermitian systems and have found potential applications in areas such as sensing~\cite{wiersig2020review}, lasing~\cite{peng2016chiral} and mode conversion~\cite{doppler2016dynamically}. 2D band structures may exhibit exceptional points within the Brillouin zone. These exceptional points have to appear in pairs, and they are connected by branch cuts in the energy band structure~\cite{miri2019exceptional}. A consequence of this branch-cut structure is that encircling an exceptional point leads to a permutation of eigenenergies~\cite{doppler2016dynamically}. Therefore, there is a direct connection between the exceptional points and the braids~\cite{artin1947theory,murasugi1999study} of eigenenergies as the momentum varies along various loops within the first Brillouin zone. Other aspects of the connection between exceptional points and eigenenergy braids have been discussed in Ref.~\cite{wojcik2022eigenvalue,hu2022knot,konig2022braid,hu2021knots,guo2022exceptional,zhang2022symmetry,li2022nonhermitian}.

Eigenenergy braiding also has important connections to point-gap topology, which is another important topological feature of non-Hermitian band structures~\cite{kawabata2019symmetry,bergholtz2021exceptional,gong2018topological}. 
In a 1D system, point-gap topology refers to the winding of the energy eigenenergies in the complex energy plane with respect to a reference energy, as the wavevector varies across the 1D Brillouin zone~\cite{bergholtz2021exceptional,gong2018topological}. A non-trivial winding number implies a non-trivial point-gap topology. A key consequence of a non-trivial point-gap topology in the band structure is the non-Hermitian skin effect. Both the point-gap topology and the non-Hermitian skin effect have been extensively discussed in the literature~\cite{borgnia2020nonhermitian,bergholtz2021exceptional,gong2018topological,okuma2020topological,yao2018edge,lin2023topological}.  In this paper, we will discuss some of the connections between the transition in braiding behaviors and the changes in the point-gap topology. 

Eigenenergy braids in 1D band structures have been experimentally demonstrated using synthetic frequency dimension in a system consisting of ring resonators~\cite{wang2021generating}. More generally, eigenenergy braids formed by varying certain parameters of the systems other than momentum have been experimentally studied in systems including cavity optomechanics~\cite{patil2022measuring}, trapped ions~\cite{cao2023probing}, mechanical oscillators~\cite{zhang2022symmetryprotected}, and acoustics~\cite{zhang2023observation,zhang2022experimental,li2022nonhermitian}. Most previous works for eigenenergy braids in energy band structures relies upon the use of tight-binding models. In this paper, we study the behaviors of eigenenergy braids in the band structure of 2D photonic crystals. 2D photonic crystal represents an experimentally accessible and technologically relevant platform for the study of band theory~\cite{joannopoulos1997photonic}. Moreover, the photonic crystal system differs from the experimental platforms as discussed above in that its band structures are usually not well described by standard tight-binding models~\cite{lidorikis1998tight}. Our study here on photonic crystals therefore highlights general aspects of the eigenenergy braids that depend on the symmetry of the system and the geometry of Brillouin zone only. These symmetries are well studied using photonic crystals~\cite{figotin2001nonreciprocal,sakoda2012proof,vaidya2023topological,alagappan2008symmetries} and do not rely upon some of the specific features of commonly used  tight-binding models, such as the presence of only a finite number of bands or the limited spatial range of coupling. We also note that more complicated, multiband structures with multiple exceptional points at various energies and momenta naturally arise in photonic crystal systems ~\cite{wang2022nonhermitian}. The study of photonic crystal band structure therefore enriches the understanding of braiding behaviors in the band theory, by providing an experimental platform to achieve complex symmetry and conjugacy classes of eigenenergy braiding~\cite{ryu2023exceptional,yang2023homotopy} that can only be observed in multiband models.

While nearly all studies on eigenenergy braiding in 2D systems focus on contractible loops encircling exceptional points~\cite{konig2022braid,wojcik2022eigenvalue,hu2022knot,guo2022exceptional}, here we focus on the eigenenergy braids on straight-line slices $k_y=m k_x + \phi$ for rational gradient $m$ and offset $\phi$ in the first Brillouin zone of a 2D system. Below we refer to these straight-line slices simply as slices. Note that these slices are also closed loops due to the periodicity of Brillouin zone. Eigenenergy braids on these Brillouin zone slices may be easier to probe experimentally than those on contractible loops around exceptional point. Point-gap topology for bands on these slices are also directly related to the geometry--dependent skin effect~\cite{fang2022geometry,zhang2021universal}. We show that the conjugacy class of eigenenergy braids on slices can only undergo a phase transition at exceptional points and we discuss the connection of such transition to change in features of the point-gap topology. 

The rest of the paper is organized as follows. In Sec.~\ref{sec_theory}, we review the relevant aspects of braid theory and how it applies to 1D and 2D non-Hermitian band structures. We describe properties of the eigenenergy braids and the connections to exceptional points in 2D systems for Brillouin zone slices. We then show the connection between the eigenenergy braids and point-gap topology. In Sec.~\ref{sec_results}, we provide examples of the theory using numerical calculations of band structures for 2D reciprocal and non-reciprocal photonic crystals. We conclude in Sec.~\ref{sec_conc}.

\section{Theory}
\label{sec_theory}
\begin{figure}
\vspace{10 pt}
\centering
\includegraphics[width= 0.45 \textwidth]{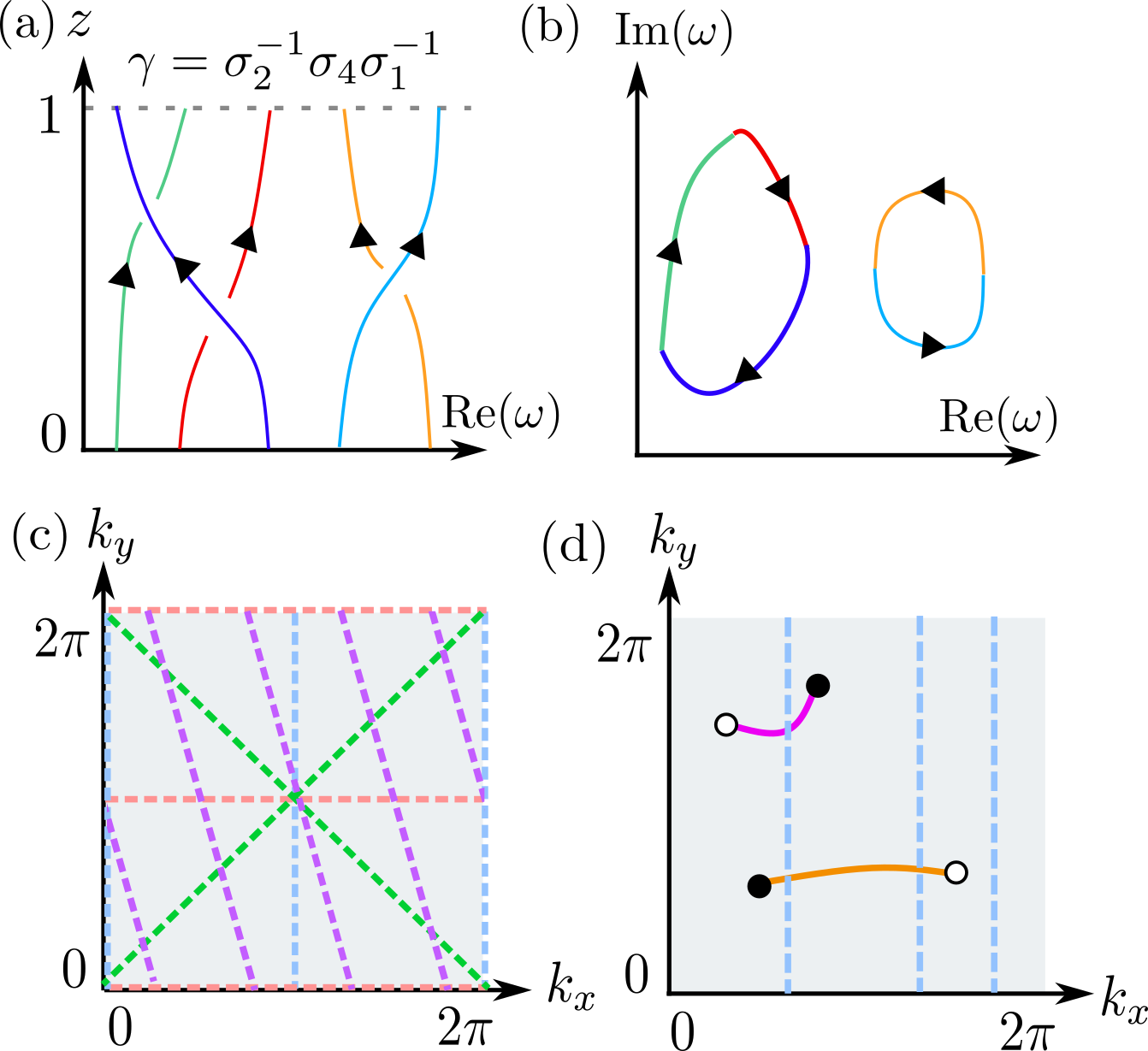}
\caption{(a) Example of a projection of an eigenenergy braid to the 2D $(\Re(\omega),z)$ plane. (b) Example of a point-gap loop corresponding to eigenenergy braid in (a). (c) Slices in the 2D Brillouin zone. (d) Three Brillouin zone slices that correspond to inequivalent braids. The filled and unfilled dots are opposite charged exceptional points and the magenta and orange lines are branch cuts.}
\label{setup}
\end{figure}
\subsection{Basics of braid theory}
We first discuss some basics of braid theory. An $N$-strand braid consists of $N$ strands in the 3D space $(\Re(\omega), \Im(\omega), z) \in \mathbb{C} \times [0,1)$. The $N$ strands don't touch, the set of endpoints at $z=0$ and 1 are fixed and each strand intersects each plane $\mathbb{C} \times \{z\}$ at only one point~\cite{gonzalez2011basic}. Another way to think of braids is as the time history of the motion of non-colliding particles in 2D space $(\Re(\omega), \Im(\omega)) \in \mathbb{C}$ where $z$ plays the role of time. Two braids $\gamma$ and $\gamma '$ are equivalent if they can be deformed into each other without breaking or intersecting any of the strands while keeping the endpoints fixed. The notion of equivalence in braids can be defined more precisely as an ambient isotopy~\cite{murasugi1999study}. 

A visualization of braids can be obtained by projecting the 3D braid in $\mathbb{C} \times[0,1)$ to the plane $\mathbb{R} \times[0,1)$. Typically, the projection from $(\Re(\omega), \Im(\omega), z)$ onto $(\Re(\omega), z)$ and $( \Im(\omega), z)$ are convenient choices. Fig.~\ref{setup}(a) depicts an example of a braid projected to the $(\Re(\omega),z)$ plane. A crossing is depicted by the strand underneath having white spaces just before and after the intersection. For an $N$-braid, we can introduce a set of generators $\{ \sigma_1 ,\sigma_2 , \hdots ,\sigma_{N-1} \}$ (called Artin generators~\cite{artin1947theory}) to describe its projection. Typically, we number the strands 1 to $N$ chronologically along the $\mathbb{R}$ axis. Here, we define $\sigma_i$ to be an over-crossing from the $i$-th to the $i+1$-th strand, and $\sigma_i^{-1}$ defines an under-crossing from the $i$-th to the $i+1$-th strand. $\sigma_i^{-1}$ is the inverse operation of $\sigma_i$~\cite{murasugi1999study}. A braid can be described by multiplying these generators to form an Artin braid word. In Fig.~\ref{setup}(a), the braid operations are read from bottom to top, so this braid is represented by the braid word $\gamma=\sigma_2^{-1} \sigma_4 \sigma_1^{-1}$ where the order of multiplication in the braid word is left to right. These generators also have the following relations:
\begin{align} 
\sigma_i \sigma_j & =\sigma_j \sigma_i,|i-j|>1 \label{braidrel1}\\ \sigma_i \sigma_j \sigma_i & =\sigma_j \sigma_i \sigma_j,|i-j|=1.
\label{braidrel2}
\end{align}
 

\subsection{Braids in 2D non-Hermitian band structures}
We now discuss how braid theory can be applied to non-Hermitian band structures.
A single energy band $\omega(k)$ in one dimension can be regarded as a strand of a braid in the $(\Re(\omega), \Im(\omega), k)$-space as the momentum $k$ is restricted to the first Brillouin zone $[0, 2\pi)$. We define two bands to be separable if their strands do not intersect, i.e. there is no degeneracy at any $k$. Moreover, as the momenta at the two ends of the Brillouin zone are equivalent, the set of eigenenergies at the start and end-points for the braid at $k=0$ and $k=2\pi$ must be the same~\cite{wang2021generating,wojcik2020homotopy}. Thus, one can describe the bands in terms of knots or links~\cite{hu2021knots}. However, the knot contains less information than the braid as many different braids may map to the same knot. In an $N$-band Hamiltonian if there are $n<N$ bands that are separable from the rest of the bands, we can focus on studying only the $n$ strands corresponding to these $n$ bands. 

Braid theory can also be applied to 2D non-Hermitian band structures. For this purpose, we consider a 1D closed path in the 2D Brillouin zone. Let us consider a 2D periodic non-Hermitian system described by an $N$-band non-Hermitian Hamiltonian with band structure $\omega(\mathbf{k})=\omega\left(k_x, k_y\right).$ The first Brillouin zone is a 2-torus $\mathbb{T}^2 =\{ (k_x,k_y) \mid k_x, k_y \in [0,2\pi) \}$. Each point $\mathbf{k} \in \mathbb{T}^2$ maps to $N$ complex numbers $\omega_1 (\mathbf{k}), \omega_2(\mathbf{k}), ...,\omega_N(\mathbf{k})$. Let us define $\Delta$ as the exceptional locus, which is the set of all wave vectors that map to degeneracies in $\omega(\mathbf{k})$. If we consider a closed loop $l$ in the Brillouin zone that avoids the exceptional locus (mathematically, $l(z):[0,1)\rightarrow \mathbb{T}^2-\Delta$), then eigenenergies on the loop $l$, i.e. $\omega(l(z))$ forms a well-defined $N$-strand braid in the 3D space defined by $(\Re(\omega), \Im(\omega), z) \in \mathbb{C} \times [0,1)$.

For eigenenergy braids in 2D systems, nearly all papers have focused on loops contractible in $\mathbb{T}^2$ around an exceptional point~\cite{wojcik2022eigenvalue,hu2021knots,hu2022knot,zhang2022symmetry,guo2022exceptional,konig2022braid}. The study of these contractible loops have led to interesting results on the properties of the exceptional points such as doubling theorems~\cite{hu2021knots,yang2021fermion,konig2022braid,wojcik2020homotopy}, non-Abelian properties~\cite{pap2018nonabelian,guo2022exceptional,wojcik2022eigenvalue} and the topological stability~\cite{wojcik2022eigenvalue,konig2022braid,yang2021fermion}. 

In contrast to these prior works, in this paper, we focus on a specific subset of closed loops in the Brillouin zone which we call \textit{slices} in the 2D Brillouin zone. These slices follow the path $k_y=m k_x + \phi$ where $m$ is the gradient and $\phi$ is the offset (Fig.~\ref{setup}(c)). Here, we choose $m$ to be rational so that the slice forms a closed loop due to the periodicity of the Brillouin zone. These loops are fundamentally different to the loops encircling an exceptional point described above, as they are non-contractible loops in $\mathbb{T}^2$. Thus, there is no longer a notion of the number of exceptional point enclosed by these slices. Nevertheless, the eigenenergy braids on slices also contain a lot of information and may be easier to measure than contractible loops around exceptional point. For example, for a photonic crystal one may directly probe certain slices by measuring the eigenenergies at a constant angle of incidence. In the next subsections, we will see that the eigenenergy braids on slices can only transition at exceptional point and are constrained by symmetry. We will also see that they have connections to point-gap topology.

\subsection{Symmetry consequences of eigenenergy braids on slices in the 2D Brillouin zone}
\label{sec_theory_symmetry}
The eigenenergies $\omega(\mathbf{k})$ traced out by a closed loop $ l(z) \in \mathbb{T}^2-\Delta$ forms a reduced  1D complex band structure $\omega (l(z))$ when plotted against the loop trajectory parameter $z$ in $[0,1)$. We call this reduced 1D band structure symmetric if there exists a particular choice of parameter $z$ such that $\omega(l(z)) = \omega(l(1-z))$. An eigenenergy braid that is symmetric in this way is trivial. If the first half of the braid contains braid operations defined by a braid word $\gamma$, then the second half, by symmetry, must undo these operations and thus must be $\gamma^{-1}$. The total braid is  $\gamma \gamma^{-1}=1$, where 1 is the trivial braid. Note that a trivial eigenenergy braid does not mean that there are no overcrossings or undercrossings. It has also been shown that symmetric 1D bands lead to trivial point-gap topology~\cite{zhong2021nontrivial}.

Let us now consider 2D reciprocal systems, which satisfy Lorentz reciprocity in electromagnetism~\cite{haus1984waves,asadchy2020tutorial}. The Lorentz reciprocity applies to systems described by symmetric permittivity and permeability tensor. For non-Hermitian systems, it differs from time-reversal symmetry. Reciprocity provides the constraint that 
\begin{equation}
\omega\left(k_x, k_y\right)=\omega\left(-k_x,-k_y\right) .
\label{reciprocity}
\end{equation}
We emphasize that Eq.~\eqref{reciprocity} can be derived using Lorentz reciprocity alone without the need from either spatial inversion or time-reversal symmetry. On the other hand, Eq.~\eqref{reciprocity} does not imply reciprocity. Eq.~\eqref{reciprocity} can be satisfied in non-reciprocal systems with either spatial inversion or time-reversal symmetry~\cite{buddhiraju2020nonreciprocal,figotin2001nonreciprocal}. 

Let us plot the Brillouin zone, $k_x,k_y \in [0,2\pi)$ as a square as in Fig.~\ref{setup}(c)-(d). Let us call vertical slices at a constant $k_x$ (sweeping from $k_y=0$ to $2\pi$) a $k_x$-slice. For a $k_x$-slice, the loop can be parametrized by $k_y$. Similarly, let us call horizontal slices at a constant $k_y$ (sweeping from $k_x=0$ to $2\pi$) a $k_y$-slice. A $k_y$-slice can be parametrized by $k_x$. Reciprocity implies that certain high-symmetry slices in the 2D Brillouin zone yield symmetric 1D band structures. The $k_x$-slices for $k_x=n\pi$ with $n$ being an integer (light blue lines in Fig.~\ref{setup}(c)) yield symmetric band structures since from reciprocity and periodicity of the Brillouin zone, $\omega(n\pi,k_y)=\omega(n\pi,-k_y)=\omega(n\pi,2\pi -k_y)$. We can apply a similar argument for $k_y$-slices at $k_y=n\pi$ (red lines in Fig.~\ref{setup}(c)). Reciprocity also implies that $\omega(k_x,k_y)=\omega(-k_x,-k_y)=\omega(2\pi-k_x,2\pi-k_y)$. This implies any reduced 1D bandstructure on any diagonal lines as defined by $k_y = mk_x$ (green or purple lines in Fig.~\ref{setup}(c)) is symmetric. All these slices yield trivial eigenenergy braids and trivial point-gap topology.

Geometric mirror symmetry can also lead to trivial eigenenergy braids. For example, if a system has mirror symmetry about the $k_y=0$ axis then $\omega(k_x,k_y)=\omega(k_x,-k_y)=\omega(k_x,2\pi-k_y)$ which means all eigenenergy braids on $k_x$ slices are trivial. Similarly, if there is mirror symmetry about the  $k_x=0$ axis, then eigenenergy braids on all $k_y$-slices are symmetrical and hence trivial. 

\subsection{Exceptional points are phase transitions for eigenenergy braids on Brillouin zone slices}
We now discuss the connection between eigenenergy braids on slices with exceptional points in the Brillouin zone. For simplicity, in this paper we restrict our focus on a 2D band structure with only order-2 exceptional points, where the order of an exceptional point indicates how many bands become degenerate at the exceptional point. We note that higher-order exceptional points are topologically unstable in 2D~\cite{yang2021fermion}. 

Order-2 exceptional points have been shown to come in pairs of opposite charges~\cite{yang2021fermion}. In 2D momentum space, these pairs of exceptional point are connected by a branch cut where the real parts of the eigenenergies become degenerate. Such a branch cut is called a bulk Fermi arc~\cite{zhou2018observation,tlusty2021exceptional,mitscherling2021nonhermitian,kozii2017nonhermitian}, where the real parts of the eigenenergies become degenerate. This is not to be confused with Fermi arcs in 3D semimetals, which are a surface rather than a bulk phenomenon~\cite{armitage2018weyl}. The exceptional points are also connected by an imaginary Fermi arc~\cite{kroll2022annihilation,bergholtz2021exceptional,yang2023realization, mandal2021symmetry} where the imaginary parts of the eigenenergies become degenerate. 

On a trajectory in momentum space upon crossing a branch cut, the corresponding eigenenergies undergo a braid operation as the crossing results in a permutation of eigenenergies from one solution branch to another. It can be shown using based loops and homotopy theory that eigenenergy braids on two parallel slices are in the same conjugacy class of the braid group $B_N$ if there is no exceptional point between the slices (see~\ref{supp_parallelconjugate}). Here, $N$ is the number of bands. Thus, we can only get inequivalent conjugacy classes for the braids of two different but parallel slices if there is one or more exceptional points between the two Brillouin zone slices. Note that there is some subtlety in the notion of ``between" for two slices on a torus. If we pick one of the enclosed areas between the two slices as $A$, then the complement of $A$ can also be considered an enclosed area between the two slices. 

As an example, in Fig.~\ref{setup}(d) we show two branch cuts drawn in magenta and orange connecting filled and unfilled dots which represent exceptional points with opposite charges. Crossing the magenta and orange branch cuts corresponds to a braid operation of $\sigma_m, \sigma_o$ respectively. We also show three slices in dotted blue lines. Each crosses two, one and zero branch cuts respectively and the braids on each slice have different braid words ($\sigma_o \sigma_m^{-1}, \sigma_o, 1$) respectively. The transitions between inequivalent conjugacy classes of braids happen when the slice intersects with an exceptional point.  

\subsection{Connections between eigenenergy braids and point-gap topology}

In addition to eigenenergy braids, non-Hermitian Hamiltonians may also exhibit point-gap topology. A 1D band structure, or a reduced 1D band structure for a 2D system, may exhibit non-trivial point-gap topology if the trajectory of the eigenenergies encloses an area in the complex energy plane. When a 2D system is truncated in the direction perpendicular to the slice direction with open boundary conditions, the system  exhibits the non-Hermitian skin effect if the energy bands on the slice exhibit non-trivial point-gap topology. This is called the geometry--dependent skin effect~\cite{zhang2021universal,fang2022geometry}. 

We now discuss the relation between point-gap topology and eigenenergy braiding. We first note they refer to different topological properties. For example, a one-band model only has trivial eigenenergy braid, but can nevertheless have non-trivial point-gap topology with different point-gap winding numbers~\cite{zhang2021acoustic,wang2021generating}. Thus, eigenenergy braid invariants and point-gap winding numbers describe different physical concepts.

There are however, connections between these two concepts. We first illustrate this connection in 1D systems. In a one-band model, starting from one end of the Brillouin zone, as the momentum evolves across the Brillouin zone to reach the other end, the eigenenergy necessarily returns to the same starting value. Thus, in the complex energy plane, upon having the momentum going through the first Brillouin zone once, the eigenenergy forms a closed loop. This loop can be thought of as the projection of the energy band into the complex energy plane $(\Re(\omega), \Im(\omega))$. Below, we refer to a closed loop in the complex energy plane as a point-gap loop. A non-trivial point-gap loop is one that encloses an area in the complex energy plane.

In an $N$-band model in 1D, suppose we similarly start with a set of the eigenenergies at one end of the first Brillouin zone. As the momentum evolves across the Brillouin zone to the other end, each individual eigenenergy may not return to its starting value. But the set of eigenenergies must remain the same at the two ends.  Consequently, the set of eigenenergies may go through a permutation as the momentum traverses across the first Brillouin zone. Suppose a non-trivial permutation does occur, for a certain starting eigenenergy then, it will require multiple traversal of the momentum across the first Brillouin zone for the eigenenergy to return to its initial starting point. The number of traversal required for the eigenenergy to return to its initial starting value is the number of bands in a complete point-gap loop. Thus, the number of bands in a point-gap loop is directly related to the permutation behavior or braids of the band.

Since the permutation behavior is characterized by the symmetry group $S_N$, one can show that the aspects of the point-gap loop as discussed above can be determined by analyzing the conjugacy class of the symmetry group $S_N$~\cite{ryu2023exceptional}. A conjugacy class in $S_N$ can be succinctly denoted as~\cite{ludwig2012symmetries}:
\begin{equation}
\prod q^{n_q}=1^{n_1} 2^{n_2} \ldots N^{n_N}
\label{conjSn}
\end{equation}
Here $q$ is the length of a cycle, and $n_q$ is the number of such $q$-cycles in a representative permutation in the conjugacy class. We also have that $\sum_q q n_q=N$. The product in Eq.~\eqref{conjSn} is a formal one, i.e. it is not a numerical product. The conjugacy class determines the aspects of the point-gap loop as mentioned above. For example, a $N=5$ band structure as described by the conjugacy class of $3^1 2^1$ have two point-gap loops, one consisting of 3 bands and the other 2 bands. 

There is a natural homomorphism from the braid group $B_N$ to the symmetry group $S_N$~\cite{murasugi1999study}. Therefore, by analyzing the braid as formed by the band structure, and by identifying the conjugacy class in $B_N$ that such a braid belongs to, one can obtain the corresponding conjugacy class in $S_N$, and hence the information on the aspects of the point-gap loops as well. An example of an eigenvalue braid and its corresponding point-gap
loop for this conjugacy class is shown in Fig.~\ref{setup}(a) and (b).

In a 2D system with $N>1$ bands, the energy bands form braids along 1D closed paths in the first Brillouin zone such as the slices discussed above. As noted previously, braids on different parallel slices may belong to different conjugacy classes in $B_{N}$ if there are exceptional points between them. As one varies the locations of the parallel slices, the transition between conjugacy classes in $B_{N}$ occur at the exceptional points. From the discussion above, these transition in braid behaviors naturally leads to a transition in the topology of the point-gap loops as well. Here, the topological property is not the point-gap winding number, but the number of bands in point-gap loops. This property is relevant for clarifying the topological invariants in multiband systems~\cite{nehra2022topology}.

\section{Numerical studies of braiding in photonic crystal band structures}
\label{sec_results}
\begin{figure}
\centering
\includegraphics[width= 0.4\textwidth]{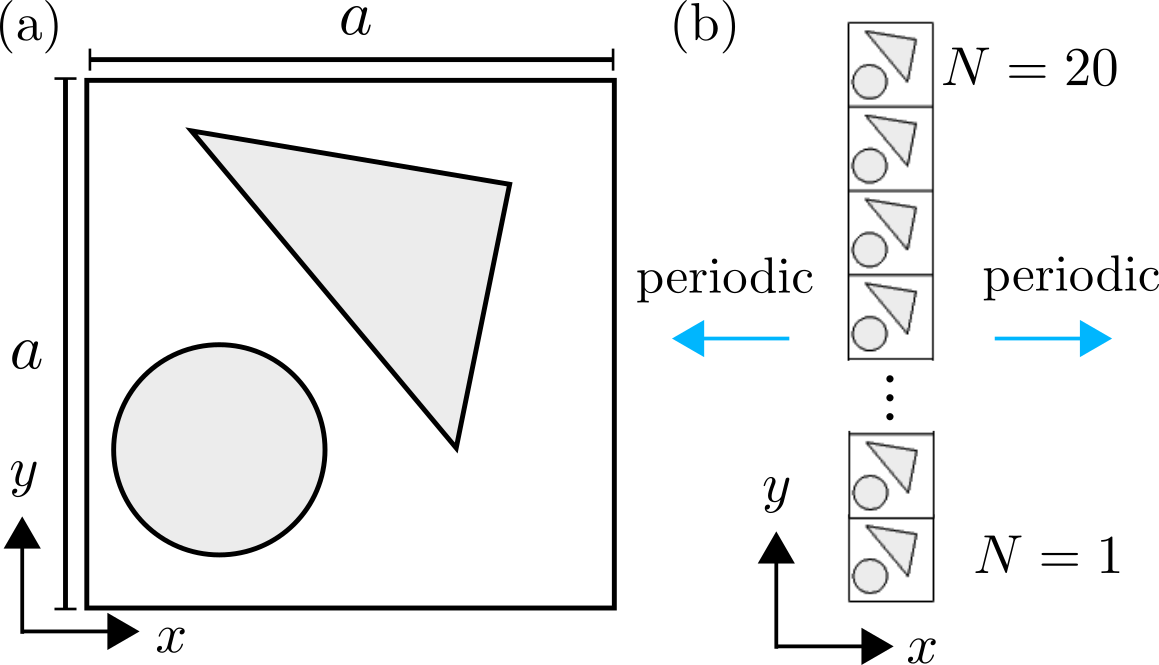}
\caption{(a) Unit cell of photonic crystal used in numerical studies. Each unit cell consists of a dielectric circular rod  that has radius $0.2a$ and is centred at $(0.25,0.3)a$ (let the bottom left corner have the coordinates (0,0)). There is also an irregular triangular dielectric rod whose vertices are at the points $(0,2, 0.9)a, (0.8,0.8)a, (0.7,0.3)a$. The circular and triangular rods are surrounded by air with refractive index $n=1$. (b) Vertical stripe of $N=20$ unit cells which has periodic boundary conditions along $x$. }
\label{phc_setup}
\end{figure}
\begin{figure*}
\centering
\includegraphics[width= 0.65\textwidth]{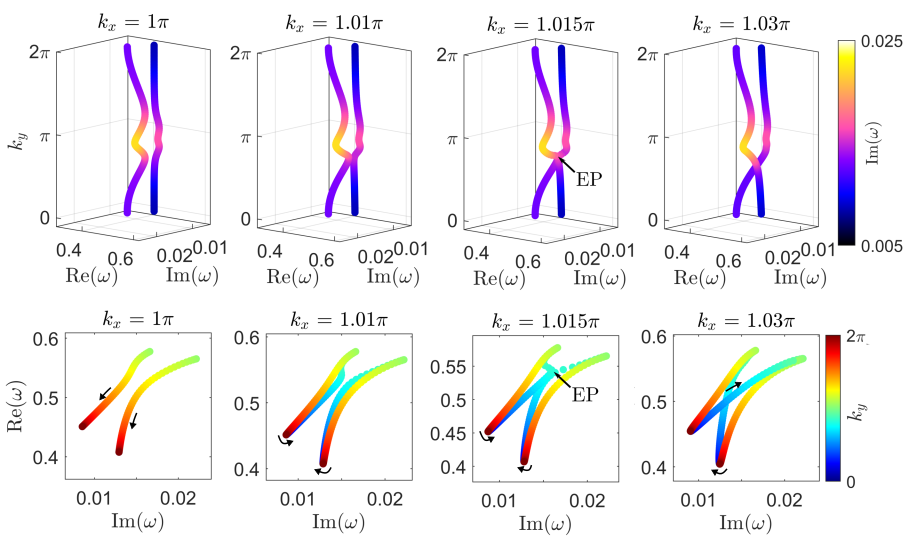}
\caption{Example of eigenenergy braid trajectory for a reciprocal photonic crystal as shown in Fig.~\ref{phc_setup}(a) with $\varepsilon_d = 3-0.6i$ and $\varepsilon_a=0$ at slices of the Brillouin zone along $k_x = \pi, 1.01 \pi, 1.015 \pi, 1.03 \pi$. In the first row, we plot these braids in the three dimensional space of complex $\omega$ and $k_y$. In the second row, we plot the projection of each of these braids onto the complex energy plane. Here all frequencies are in the unit of $2\pi c/a$. We see that an exceptional point is crossed at approximately $k_x=1.015\pi$ which corresponds to a conjugacy class transition in the braids.}
\label{reciprocal}
\end{figure*}
Motivated by the theory as discussed in the previous section, in this section we undertake a numerical study of eigenenergy braidings in the band structure of 2D photonic crystals. Our aims are: (1) to show that one can observe eigenenergy braidings along slices; (2) to numerically demonstrate the transition of between inequivalent conjugacy classes of braids on different parallel slices and to connect such transition with exceptional points; (3) to illustrate the change in aspects of point-gap topology as induced by the transition of the braiding behaviors. 

For our aims, we consider a 2D square-lattice photonic crystal with lattice periodicity $a$ where the unit cell is shown in Fig.~\ref{phc_setup}(a). We note that such a structure does not possess any mirror or inversion symmetry, as motivated by the discussion in Sec.~\ref{sec_theory_symmetry}. 

We assume that the permittivity in the dielectric rods have the form:
\begin{equation}
\varepsilon=\left[\begin{array}{ccc}
\varepsilon_d & -i \varepsilon_a & 0 \\
i \varepsilon_a & \varepsilon_d & 0 \\
0 & 0 & \varepsilon_d
\end{array}\right].
\end{equation}
When $\varepsilon_a = 0$, the material is a regular dielectric with a permittivity $\varepsilon_d$ and is reciprocal. When $\varepsilon_a \ne 0$, the material is a magneto-optical material with its magnetization along the $z$-direction, and it is non-reciprocal. We consider a TM polarized wave propagating in the $x y$ plane with a magnetic field in the $z$ direction and an electric field in the $x y$ plane. Throughout the paper, we use the $e^{i \omega t}$ convention. Thus, an imaginary component of $\varepsilon_d$ that is negative means that the material is lossy and the structure is non-Hermitian. We numerically determine the 2D photonic band structure $ \omega(k_x, k_y)$ of the photonic crystal system shown using COMSOL Multiphysics~\footnote{\url{https://github.com/fancompute/Point-gap-Topology-Braids-PhC}}, which employs finite-element methods to numerically solve Maxwell's equations. The eigenenergies of the energy are in general complex. From the 2D band structure, we then determine the 1D band structure along various slices.

\begin{figure}
\centering\includegraphics[width=0.5\textwidth]{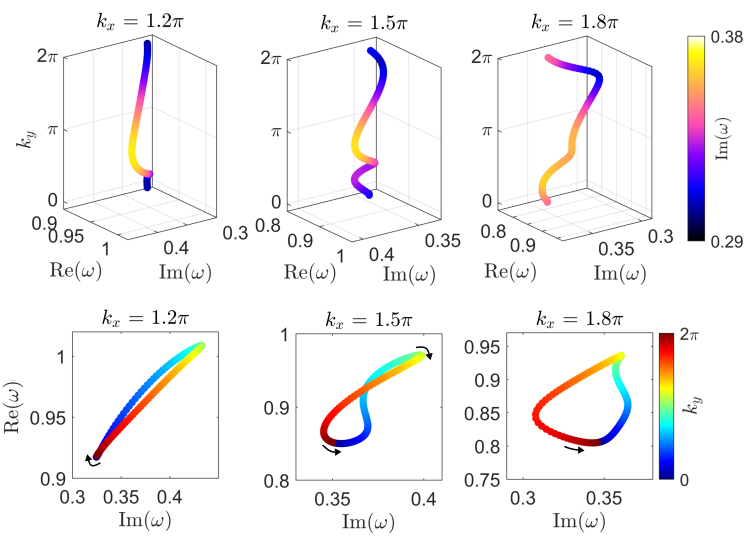}
\caption{ Example of one-band trivial eigenenergy braid that corresponds to a non-trivial point-gap topology for a nonreciprocal photonic crystal as shown in Fig.~\ref{phc_setup} with $\varepsilon_d = 2-i$ and $\varepsilon_a=3$.}
\label{figureeight}
\end{figure}
In Fig.~\ref{reciprocal}, we study a case of a reciprocal system where $\varepsilon_d = 3-0.6i$ and $\varepsilon_a=0$. In the top row, we plot eigenenergy braids for two bands in $\omega(k_x,k_y)$ for this system for $k_x$-slices at $k_x=\pi, 1.01\pi, 1.015\pi$ and $1.03\pi$. In the second row of Fig.~\ref{reciprocal}, we plot the projection of the energy band to the complex energy plane for the same slices. 

On the $k_x$-slice at $k_x=\pi$ (left panels in Fig.~\ref{reciprocal}), the photonic bands exhibits a trivial braid and trivial point-gap topology, as expected from the discussion in Sec.~\ref{sec_theory_symmetry}. On the $k_x$-slice at $k_x=1.01\pi$ (middle-left panels in Fig.~\ref{reciprocal}), we move away from the high symmetry line in the Brillouin zone and we see that the projection on the complex energy plane consists of two loops with nontrivial point-gap topology. On the $k_x$-slice at $k_x=1.015\pi$ (middle-right panels in Fig.~\ref{reciprocal}), the two strands intersect at an exceptional point which is depicted by an arrow in the second panel in Fig.~\ref{reciprocal}. The eigenenergy braid is not well-defined and two point-gap loops merge at the exceptional point. On the $k_x$-slice at $k_x=1.03\pi$ (right panels in Fig.~\ref{reciprocal}), which does not cut through an exceptional point, the two bands swap eigenenergies as $k_y$ changes from 0 to $2\pi$. This is a nontrivial braid described by braid-word $\sigma_1^{-1}$ when the band is projected onto the $(\Re(\omega),k_y)$ plane. The projection on the complex energy plane is now a joint two-band point-gap loop in a figure-eight shape. 

The results in Fig.~\ref{reciprocal} represent a conjugacy class transition in the eigenenergy braids on the $k_x$ slices as we vary $k_x$ from 1.01 to $1.03 \pi$. In order to go from the trivial two-band braid at $k_x=1.01\pi$ to a nontrivial two-band braid at $k_x=1.03\pi$ via a continuous change of $k_x$, an exceptional point has to be crossed, as we see in the case of $k_x = 1.015\pi$. By examining the projection of the bands to the complex energy plane, as we do in the bottom panels of Fig.~\ref{reciprocal}, we also see that this transition of the braiding behavior at the exceptional point corresponds to the transition on aspects of the point-gap loops. As the braids transitions, the point-gap loops transitions from having two separate loops each corresponding to one band to having a single loop for the two bands. 

We next illustrate the difference between point-gap topology and eigenenergy braiding in the example of Fig.~\ref{figureeight}. We use the same unit-cell geometry in Fig.~\ref{phc_setup}(a) but with $\varepsilon_d = 2-i$ and $\varepsilon_a=3$. As the permittivity tensor is no longer symmetric, this system is nonreciprocal. We pick a region in the band structure where there is a single band that has nontrivial point-gap topology. In the left panel of Fig.~\ref{figureeight}, we plot the eigenenergy braid and its projection onto the complex energy plane for $k_x$-slice at $k_x=1.2\pi$. As it is only a single strand, the braid is trivial. However, from the projection, we see that the band forms a loop that encloses non-trivial area. As $k_y$ varies from $0$ to $2\pi$, the complex energy moves clockwise on the loop. Thus, the loop has a winding number of $W=-1$~\cite{gong2018topological} with respect to any reference energy inside the loop. As we continuously vary the the locations of the $k_x$-slice, the topology of the point gap loop can change. At $k_x=1.5\pi$ (middle panel, Fig.~\ref{figureeight}), the point gap loop has a figure-8 shape with $W=-1$ with respect to a reference energy inside the top half of the loop and $W=1$ with respect to a reference energy in the bottom half loop. At $k_x = 1.8\pi$, the loop becomes a simple loop with a winding number of $W = 1$ with respect to an energy inside the loop. In this example, we see that the system can exhibit non-trivial point gap topology, and even transitions between different point-gap topology, while the braid behavior is trivial. The example therefore highlights the difference between the concepts of point-gap topology and eigenenergy braiding.  

\begin{figure}
\centering
\includegraphics[width= 0.4\textwidth]{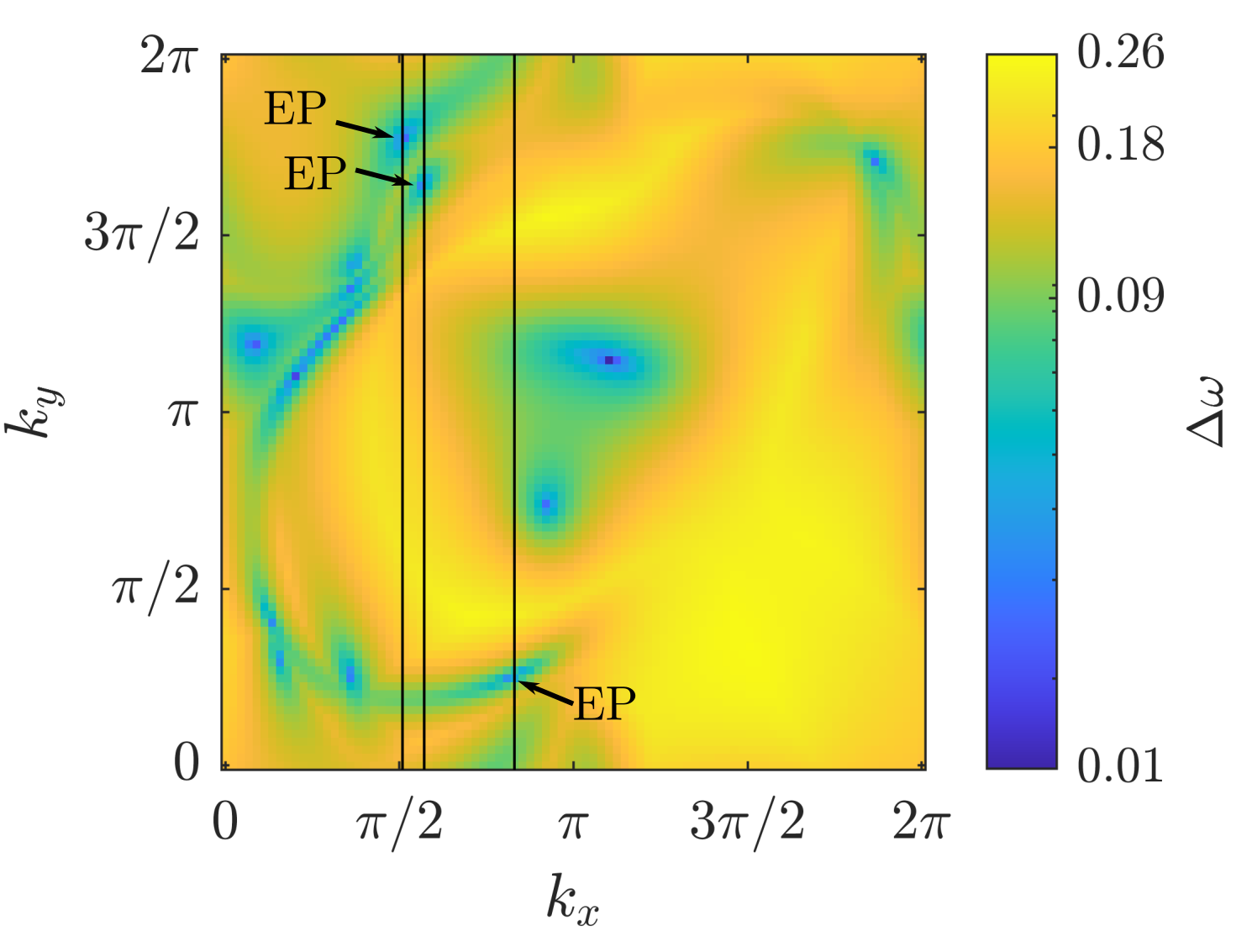}
\caption{Minimum eigenenergy distance $\Delta \omega$ (in units of $2 \pi c / a$) in logarithmic scale for a nonreciprocal photonic crystal shown in Fig.~\ref{phc_setup}(a) where $\varepsilon_d = 1-0.6i$ and $\varepsilon_a=2$ in the region $\operatorname{Re}(\omega) \in[0.6, 1.150]\cdot 2\pi c / a$ and $\operatorname{Im}(\omega) \in[0.05 , 0.55 ]\cdot 2\pi c / a$. We numerically identify 14 exceptional points which are blue dots where $\Delta \omega$ goes to zero. Three of the exceptional points are labelled and these intersect with $k_x$-slices at $k_x=0.51 \pi, 0.57\pi$ and $0.83\pi$. These $k_x$-slices have been plotted in black.}
\label{mineig}
\end{figure}

\begin{figure*}
\centering\includegraphics[width=\textwidth]{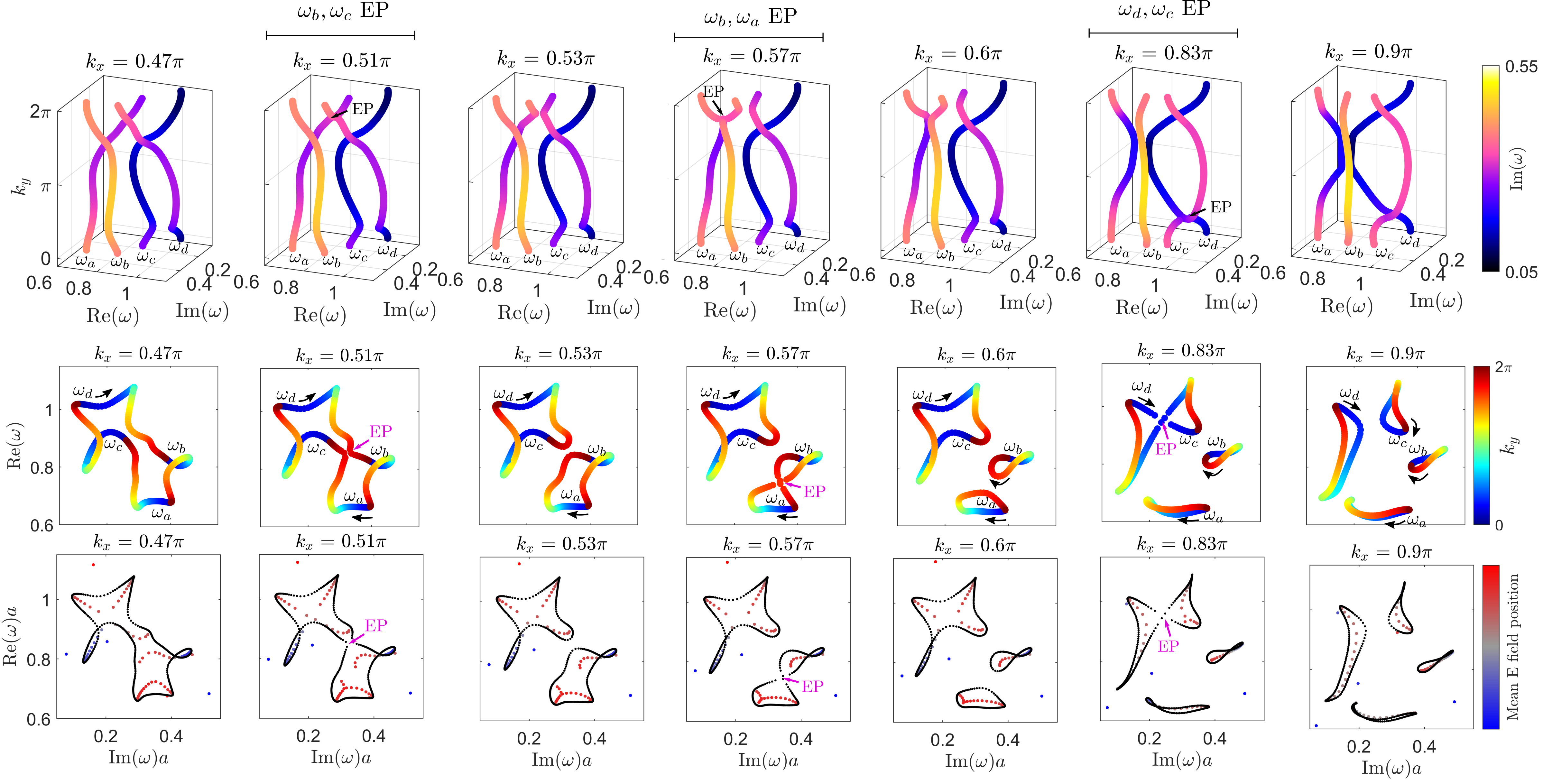}
\caption{In the top row, we have the eigenenergy braid for nonreciprocal photonic crystal setup with $\varepsilon_d = 1-0.6i$ and $\varepsilon_a=2$ on slices of the Brillouin zone along $k_x = 0.47\pi, 0.51\pi, 0.53 \pi, 0.57 \pi, 0.6 \pi, 0.63\pi , 0.8\pi, 0.9 \pi$. In the middle row, we have the projections of the eigenenergy braid on the complex energy plane for same parameters. In the bottom row, we show the complex eigenenergy projection in black and the geometry--dependent skin-effect for vertical stripe geometry for $kx$-slices in the red-grey-blue color map. Here, blue or red indicates top or bottom localization of the electric field. }\label{nonreciprocal_slices}
\end{figure*}

Finally, we show an example with more complex braiding and braiding transition behaviors involving multiple bands. For this purpose, we again consider the same geometry as in Fig.~\ref{phc_setup}(a), but with $\varepsilon_d = 1-0.6i$ and $\varepsilon_a=2$. As $\varepsilon_a \ne 0$, the system is again nonreciprocal. In this system, there are four bands with eigenenergies entirely restricted to the region $\operatorname{Re}(\omega) \in[0.6, 1.150]\cdot 2\pi c / a$ and $\operatorname{Im}(\omega) \in[0.05 , 0.55 ]\cdot 2\pi c / a$. These four bands are separable from the rest of the bands. Moreover, we also have a nontrivial boundary braid for these four bands, which is shown in~\ref{supp_basedloopphc}. This means that as the momentum moves around the boundary of the first Brillouin zone, the four bands form a non-trivial braid. Having such a non-trivial boundary braid implies the existence of exceptional points for these four bands inside the first Brillouin zone~\cite{wojcik2022eigenvalue}. 

For these four bands, we numerically identified 14 exceptional points using Fig.~\ref{mineig}. In this figure we have numerically calculated the 2D complex band structure for a 120 by 120 grid of $(k_x, k_y)$ points for the region containing the four bands. We plot the minimum eigenenergy distance $\Delta \omega$~\cite{luitz2019exceptional} between all pair combinations of the four eigenenergies at each $(k_x, k_y)$ point. The exceptional points appear when the distance between eigenenergies becomes zero, which are blue dots in Fig.~\ref{mineig}~\cite{supplemental_material}. Among the 14 exceptional points in Fig.~\ref{mineig}, three are located at $k_x=0.51 \pi, 0.57\pi$ and $0.83\pi$. These three exceptional points are labelled by arrows in Fig.~\ref{mineig}. 

In the top row of Fig.~\ref{nonreciprocal_slices}, we show the eigenenergy braids on slices $k_x=0.47\pi, 0.51\pi, 0.53\pi,0.57\pi, 0.6\pi,0.83\pi$ and $0.9\pi$. We see that the exceptional points at $k_x=0.51 \pi, 0.57\pi$ and $0.83\pi$ all correspond to conjugacy class transitions in the eigenvalue braids. In the middle row of Fig.~\ref{nonreciprocal_slices}, we show the projection of the eigenenergy braid on the complex energy plane. At $k_x=0.47\pi$, we have a four-band point-gap loop. On this slice we label the eigenenergies of the four bands at  the starting point $k_y = 0$ as $\omega_a, \omega_b, \omega_c, \omega_d$. As $k_y$ evolves from 0 to $2\pi$, these four eigenenergies go through a permutation as described by a 4-cycle $(\omega_d \omega_b \omega_a \omega_c$). At $k_x=0.51\pi$, there is an exceptional point between the bands starting at $\omega_b$ and $\omega_c$. As $k_x$ increases from below to above $0.51 \pi$, the presence of this exceptional point corresponds to a split of the four-point point-gap loop. At $k_x=0.53\pi$, we now have two two-band point-gap loops with the cycle $(\omega_d \omega_c)(\omega_a \omega_b)$. Another exceptional point occurs at $k_x=0.57\pi$, but this time with the bands starting at $\omega_a$ and $\omega_b$. As $k_x$ increases across $k_x = 0.57 \pi$, in the projection to the complex energy plane we observe a transition where the two-band cycle $(\omega_a \omega_b)$ splits into two single-band loops. As a result, at $k_x=0.6\pi$, the projection consists of one two-band loop and two single-band loops, corresponding to the permutation cycle $(\omega_d \omega_c)(\omega_a)( \omega_b)$. Finally, we have an exceptional point  at $k_x=0.83\pi$, this time between the bands starting at $\omega_d$ and $\omega_c$. In the projection, this corresponds to the splitting of the two-band loop $(\omega_d \omega_c)$ splitting. As a result, in the projection at $k_x=0.9\pi$, we now have four single-band loops, which is described by a permutation cycle $(\omega_d)( \omega_c)(\omega_a)( \omega_b)$. 

In the bottom row of Fig.~\ref{nonreciprocal_slices}, we plot the complex eigenenergies for a system as described by a vertical stripe as shown in Fig.~\ref{phc_setup}(b) in a red-grey-blue colourscale~\cite{zhong2021nontrivial}. The vertical stripe consists of 20 photonic crystal unit cells arranged along the y-axis. Outside the photonic crystal region we have air regions backed by a perfectly electric conductor boundary condition. Since the system is periodic along $x$, we can still plot the eigenspectra for a particular $k_x$-slice. As the $y$ direction now has open boundary conditions, for each $k_x$ value we expect the geometry--dependent skin effect if the corresponding projection of the complex energy band (the second row of Fig.~\ref{nonreciprocal_slices}) has nontrivial area in the complex energy plane. In our case, we indeed observe the geometry--dependent skin effect. We observe that the complex eigenenergy of the modes in the finite system lies within the area enclosed by the projection of the energy bands, as also plotted in the bottow row. For each complex eigenenergy in this system, the red-grey-blue colour-map is given by the mean position $\bar{y}$
\begin{equation}
\bar{y}=\frac{\int_0^a d y \int_0^{N a} d x|E(x, y)| y}{\int_0^a d y \int_0^{N a} d x|E(x, y)|}
\end{equation}
of the electric field amplitude of the corresponding eigenstate. Blue indicates the electric field amplitude is localized at the top of the stripe ($N=20$) whereas red indicates the electric field amplitude is localized at the bottom edge of the stripe ($N=1)$ as in Fig.~\ref{phc_setup}(b). Grey indicates that either the field is delocalized or is localized at both edges. In the bottom role of Fig.~\ref{nonreciprocal_slices}, we see that within each point gap loop there are localized states on the edges of the stripes. These localized states are localized at the bottom edge when the winding numbers is $W = -1$ and at the top edge when the winding numbers is $W = +1$.

\section{Conclusion}
\label{sec_conc}
We consider eigenenergy braids on slices in 2D Brillouin zone. We discuss the consequences of reciprocity and geometric symmetry on these slices. We show that the braids on these slices can only undergo conjugacy class transitions when the slice intersects an exceptional point. The conjugacy class transition corresponds to a change in the number of bands in a complete point-gap loop, where the point-gap loop is the projection of the eigenenergy braid onto the complex energy plane. Thus, this transition corresponds to point-gap loops merging or splitting. Our theoretical results show the connection between the conjugacy class of eigenenergy braids, point-gap topology and exceptional points. We numerically demonstrate these concepts using two-dimensional reciprocal and nonreciprocal photonic crystals, which represent a technologically-relevant platform for the exploration of non-Hermitian topology.

\vspace{20 pt}
\section*{Acknowledgements}
 JZ acknowledges useful comments from Heming Wang,  Yidong Chong and Kai Wang. This work is funded by a Simons Investigator in Physics
grant from the Simons Foundation (Grant No. 827065).

%

\setcounter{section}{0}
\setcounter{equation}{0}
\renewcommand{\thesection}{Appendix \Alph{section}}
\renewcommand{\thesubsection}{\arabic{subsection}}
\renewcommand*{\citenumfont}[1]{#1}
\renewcommand*{\bibnumfmt}[1]{[#1]}
%

%

%
\section{Eigenenergy braids on parallel slices}
\label{supp_parallelconjugate}
In this section, we show that the braids on slices of the two-dimensional Brillouin zone are conjugate if there are no exceptional points between them. 

\begin{figure}[h]
\centering
\includegraphics[scale=0.2]{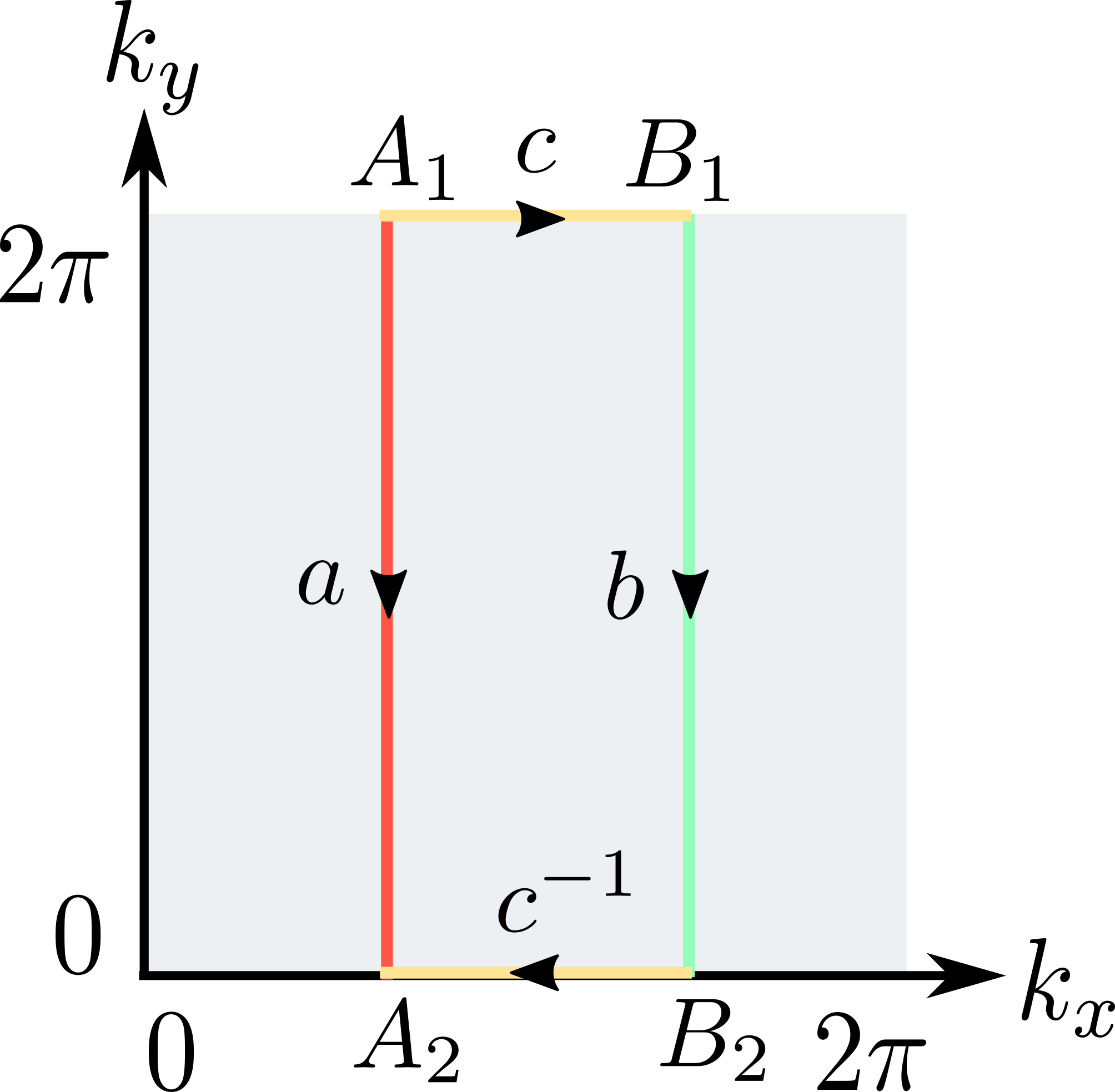}
\caption{The braids corresponding to Brillouin zone slices $b$ and $f$ are conjugate.}
\label{conjugateexample}
\end{figure}

Consider eigenenergy braids $a$ and $b$, on two parallel slices, as defined by directed line segments $A_1 A_2$ and $B_1 B_2$, respectively (Fig.~\ref{conjugateexample}). We denote as $c$ the eigenenergy braid on $A_1 B_1$. The braids on $B_2 A_2$ is then $c^{-1}$ since $A_1 B_1$ and $A_2 B_2$ are the same in the Brillouin zone. Suppose there is no exceptional point between the two slices, we then have
\begin{equation}
a=c b c^{-1}
\end{equation}
And hence $a$ and $b$ are conjugate with each other.

\section{Non-trivial boundary braid and based loop analysis}
\label{supp_basedloopphc}
\begin{figure}[h]
\centering
\includegraphics[width=0.5\textwidth]{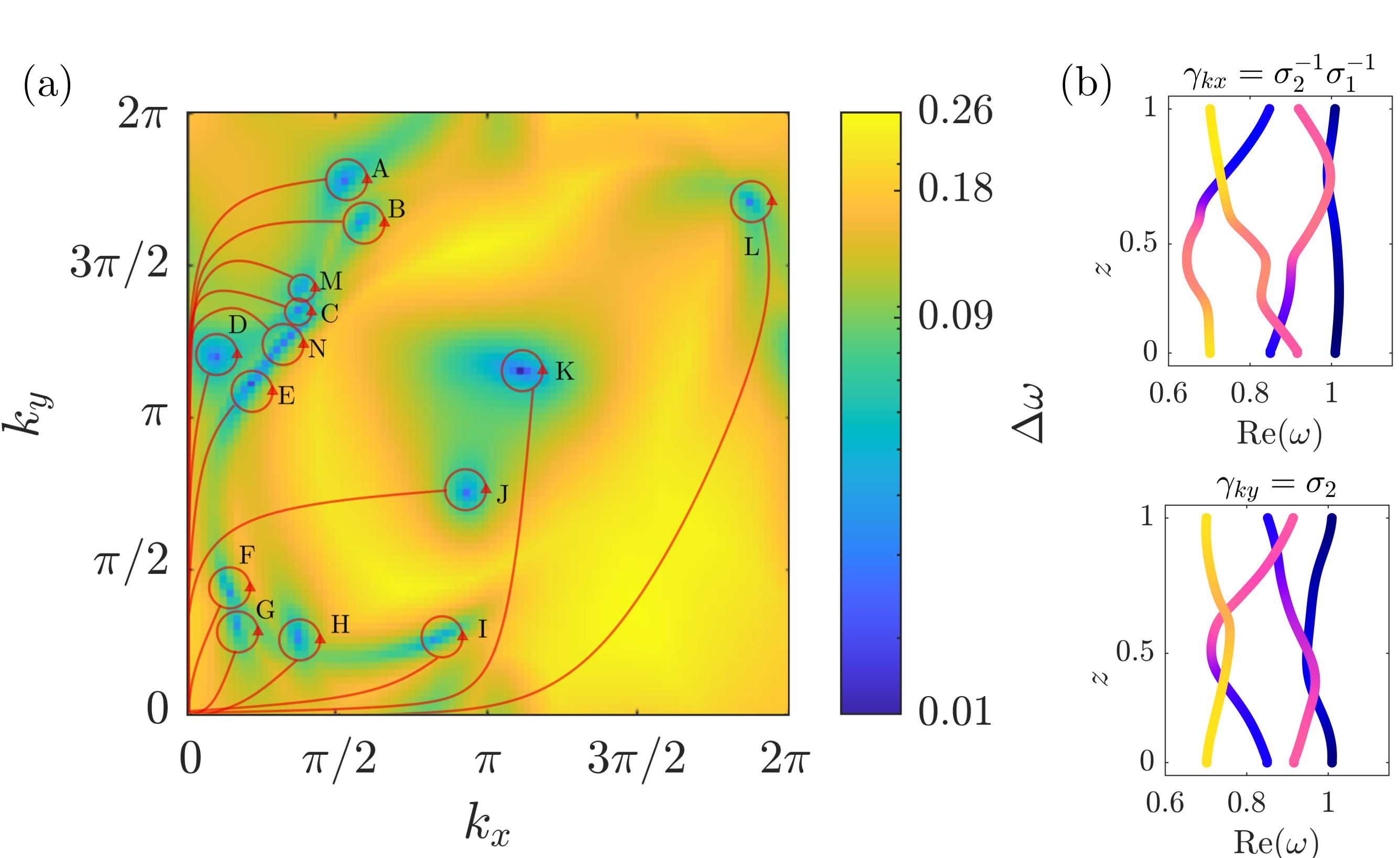}
\caption{(a) Minimum eigenenergy distance $\Delta \omega$ (in units of $2 \pi c / a$) in logarithmic scale for a nonreciprocal photonic crystal shown in Fig.~\ref{phc_setup}(a) where $\varepsilon_d = 1-0.6i$ and $\varepsilon_a=2$ in the region $\operatorname{Re}(\omega) \in[0.6, 1.150]\cdot 2\pi c / a$ and $\operatorname{Im}(\omega) \in[0.05 , 0.55 ]\cdot 2\pi c / a$. The dark blue dots are exceptional points, where the eigenenergy distance goes to zero. There are 14 exceptional points and we have drawn based loops around each, as well as the braids corresponding to these loops. (b) Nontrivial boundary braids for this set-up. The top panel is for the braids on the trajectory of $k_x \in[0,2\pi)$ where $k_y=0$. The bottom panel is for the braids on the trajectory of $k_y \in[0,2\pi)$ where $k_x=0$. $z$ labels the coordinate along the trajectory.}
\label{nonreciprocal_mineig}
\end{figure}
In the main text, we have focused on braids on slices. In the literature, there has been several works examining braids on based loops around exceptional points~\cite{wojcik2022eigenvalue,hu2022knot,konig2022braid,guo2022exceptional}. Our photonic crystal system can also be used to explicit check some of theses theoretical results.

We review the connection between eigenenergy braids and homotopic loops in the space $\mathbb{T}^2 -\Delta$, where $\Delta$ is the set of $k$ exceptional points. $\mathbb{T}^2 -\Delta$ is the $k$-punctured 2-torus. The homotopic equivalence class of based loops on $\mathbb{T}^2 -\Delta$ forms a group structure where the group product is concatenation~\cite{pap2018nonabelian}. This group is the fundamental group $\pi_1(\mathbb{T}^2-\Delta)$ of $\mathbb{T}^2-\Delta$~\cite{wojcik2022eigenvalue}.

Let us define loops around each exceptional point as 
$l_{e 1}, l_{e 2} \ldots, l_{e k}$. We assume these loops are contractible in $\mathbb{T}^2-\Delta$.  We also denote the two fundamental loops of the torus as $l_{k_x}$ and $l_{k_y}$. We assume all these loops have the same base point. Then these loops satisfy the following relation:
\begin{equation}
\underbrace{l_{e1} l_{e2} \hdots l_{ek}}_{l_{EP}}=\underbrace{l_{k_x} l_{k_y} l_{k_x}^{-1} l_{k_y}^{-1} }_{l_{BZ}}.
\label{loopeq}
\end{equation} 
where $l_{EP}$ is the product of the based, contractible loops around the exceptional points and $l_{BZ}$ is the loop around the boundary of the first Brillouin zone.  

There is a homomorphism from the group of equivalence classes of based loops to the braid group $B_N$. As a result~\cite{wojcik2022eigenvalue}:
\begin{equation}
\underbrace{\gamma_{e_1} \gamma_{e_2} \ldots \gamma_{e_k}}_{\gamma_{EP}}=\underbrace{\gamma_{k x} \gamma_{k y} \gamma_{k x}^{-1} \gamma_{k y}^{-1} }_{\gamma_{BZ}}.
\label{EPequalsBZ}
\end{equation}
where $\gamma_i$ is the eigenenergy braid on the loop $l_i$. By writing these braids using Artin's relation, we can also find that 
\begin{equation}
\mathcal{C}(\gamma_{EP})= \mathcal{C}(\gamma_{BZ}),
\label{nogotheorem}
\end{equation}
where the crossing number $\mathcal{C}(\gamma)$ of a braid $\gamma$ (also known as the exponent sum~\cite{murasugi1999study} or the writhe number~\cite{hu2021knots}) is the number of over-crossings subtracted by the number of under-crossings in the braid. This statement is equivalent to the non-Hermitian no go theorem~\cite{hu2021knots} as well as the non-Abelian sum rule~\cite{konig2022braid}. Eq.~\eqref{nogotheorem} contains less information than Eq.~\eqref{EPequalsBZ}.

We verify Eq.~\eqref{EPequalsBZ} using the same nonreciprocal photonic crystal in Fig.~\ref{nonreciprocal_slices} of the main text. Here, we consider the same four bands as discussed in the main text, which are separable from all the other bands of the system. We plot $\gamma_{kx}$ and $\gamma_{ky}$ eigenenergy braids in Fig.~\ref{nonreciprocal_mineig}(b). For all our braid projection plots, we project to the $(\Re(\omega),k)$ plane. In this projection,  we have $\gamma_{kx} = \sigma_2^{-1} \sigma_1^{-1}$ and $\gamma_{ky} = \sigma_2$. Hence, we have nontrivial $\gamma_{BZ} = \gamma_{k x} \gamma_{k y} \gamma_{k x}^{-1} \gamma_{k y}^{-1} = \sigma_1 \sigma_2^{-1}$. The existence of a non-trivial boundary braids implies the presence of exceptional point inside the Brillouin zone, in consistency with our numerical observation here. 

\begin{figure*}[t]
\centering\includegraphics[width=\textwidth]{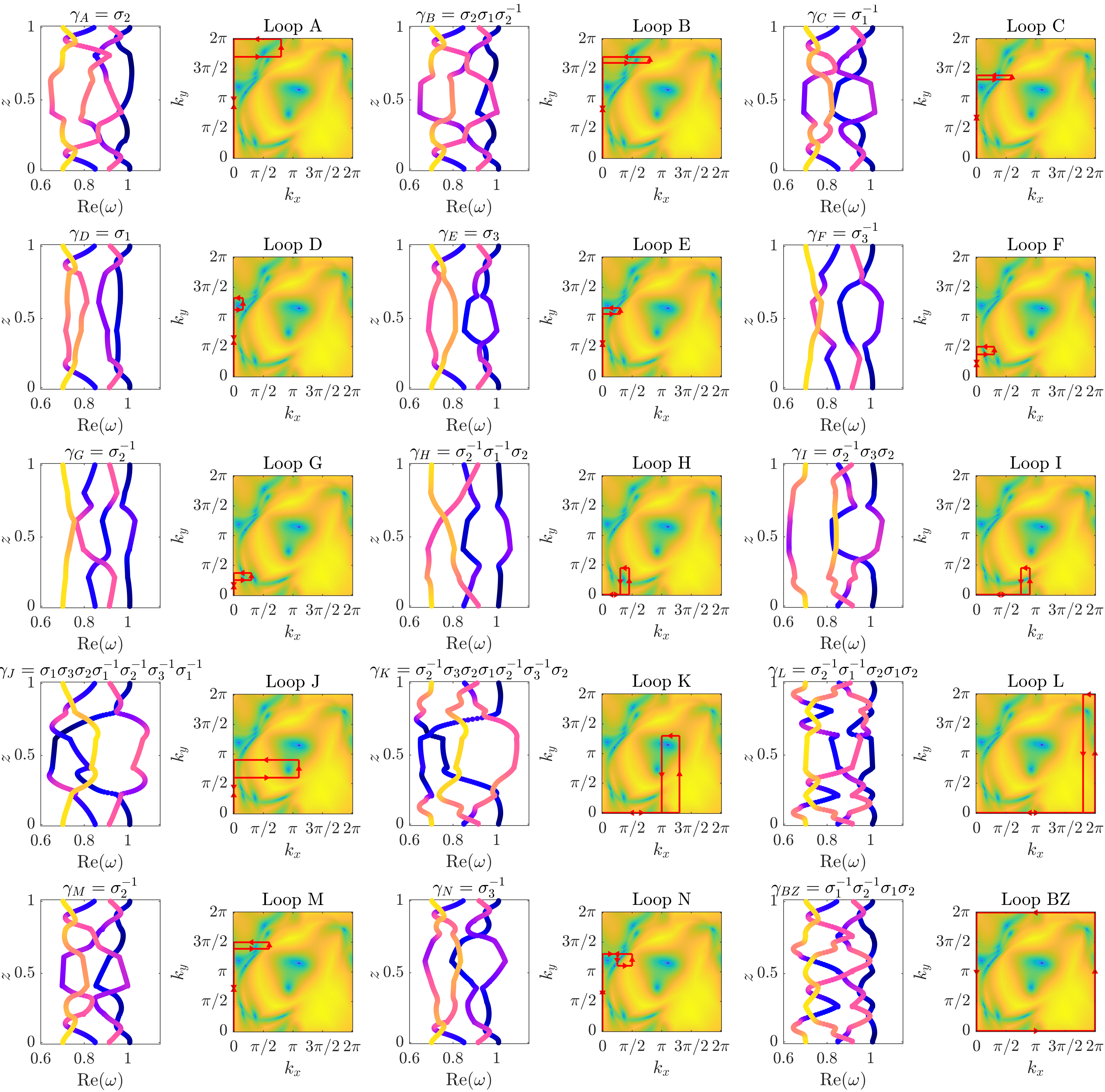}
\caption{Eigenenergy braids for the cases as described in Table \ref{braidwords}. For all plots, the base point is the origin of the Brillouin zone.}
\label{nonreciprocal_loops}
\end{figure*}

We next provide a direct check of Eq.~\eqref{EPequalsBZ}. For this purpose we first find the eigenenergy braids on based loops around the 14 exceptional points identified in Fig.~\ref{nonreciprocal_mineig}(a). In Fig.~\ref{nonreciprocal_mineig}, these exception points are labelled from $A$ to $N$, and the corresponding braids on these loops are labelled $\gamma_A$ to $\gamma_N$. We plot the eigenenergy braids on these loops in Fig.~\ref{nonreciprocal_loops}, and record them in Table~\ref{braidwords}. For this system, Eq.~\eqref{EPequalsBZ} becomes:
\begin{equation}
\gamma_L \gamma_K \gamma_I \gamma_H \gamma_G \gamma_F \gamma_J \gamma_E \gamma_D \gamma_N \gamma_C \gamma_M \gamma_B \gamma_A=\gamma_{k x} \gamma_{k y} \gamma_{k x}^{-1} \gamma_{k y}^{-1}.
\end{equation}
We can verify that this is true using the results in Table~\ref{braidwords}. In the algebraic check, note that
$ \gamma_{BZ}=\sigma_1^{-1} \sigma_2^{-1} \sigma_1 \sigma_2 = \sigma_1 \sigma_2 ^{-1} ,
 \gamma_L \gamma_k \gamma_I \gamma_k \gamma_G \gamma_F \gamma_J = \sigma_1\sigma_2^{-1} \sigma_1^{-1},
 \gamma_E \gamma_D \gamma_N = \sigma_3 \sigma_1 \sigma_3^{-1} ,
 \gamma_C \gamma_M \gamma_B \gamma_A = 1$.
From this, one can show that $\gamma_{EP}=\gamma_{BZ}$. Thus, we have verified Eq.~\eqref{EPequalsBZ} for a rather complicated system featuring multiple bands and a large number of exceptional points. The relation of Eq.~\eqref{nogotheorem} can also be checked explicitly.

\begin{table}[H]
\centering
\vspace{10 pt}
\begin{tabular}{p{2.5cm}p{4cm}p{1cm}}
\hline
braid                                                            & braid word                                                                          & crossing number \\ \hline
$\gamma_A$                                                              & $\sigma_2$                                                                          & 1                             \\
$\gamma_B$                                                              & $\sigma_2 \sigma_1 \sigma_2^{-1}$                                                   & 1                             \\
$\gamma_C$                                                              & $\sigma_1^{-1}$                                                                     & -1                            \\
$\gamma_D$                                                              & $\sigma_1$                                                                          & 1                             \\
$\gamma_E$                                                              & $\sigma_3$                                                                          & 1                             \\
$\gamma_F$                                                              & $\sigma_3^{-1}$                                                                     & -1                            \\
$\gamma_G$                                                              & $\sigma_2^{-1}$                                                                     & -1                            \\
$\gamma_H$                                                              & $\sigma_2^{-1} \sigma_1^{-1} \sigma_2 $                                             & -1                            \\
$\gamma_I$                                                              & $\sigma_2^{-1} \sigma_3 \sigma_2$                                                   & 1                             \\
$\gamma_J$                                                              & $\sigma_1 \sigma_3 \sigma_2\sigma_1^{-1} \sigma_2 ^{-1}\sigma_3^{-1} \sigma_1^{-1}$ & -1                            \\
$\gamma_K$                                                              & $\sigma_2^{-1} \sigma_3\sigma_2 \sigma_1 \sigma_2^{-1} \sigma_3^{-1} \sigma_2$      & 1                             \\
$\gamma_L$                                                              & $\sigma_2^{-1} \sigma_1^{-1} \sigma_2 \sigma_1 \sigma_2$                                                                     & 1                            \\
$\gamma_M$                                                              & $\sigma_2^{-1}$                                                                     & -1                            \\
$\gamma_N$                                                              & $\sigma_3^{-1}$                                                                     & -1                            \\ \hline
$\gamma_{kx} \gamma_{ky} \gamma_{kx}^{-1} \gamma_{ky}^{-1}$ & $\sigma_1^{-1} \sigma_2^{-1} \sigma_1 \sigma_2$                                                          & 0                             \\ \hline
\end{tabular}
\caption{Various braids in the non-reciprocal photonic crystal considered in Fig.~\ref{nonreciprocal_loops}}
\label{braidwords}
\end{table}

\nocite{apsrev41Control}
\bibliographystyle{apsrev4}
\bibliography{titleon,bib}

\end{document}